\documentclass[aps,secnumarabic,showpacs,nobibnotes,amssymb,prd]{revtex4}
\begin{document}
\title{\bf Particle Spectrum of the Supersymmetric Standard Model from the Massless Excitations 
of a Four Dimensional Superstring}
\medskip
\author{B. B. Deo and  L. Maharana\footnote{E-Mail:~ lmaharan@iopb.res.in}}
\affiliation{ Physics Department, Utkal University, Bhubaneswar-751004, India.}
\begin{abstract}
A superstring action is quantised with Neveu Schwarz(NS) and Ramond(R) boundary conditions. 
The zero mass states of the NS sector are classified as the vector gluons, W-mesons,
$B_{\mu}$-mesons and scalars containing Higgs. The fifteen zero mass fermions are obtained 
from the Ramond sector. A space time supersymmetric Hamiltonian of the Standard Model is
presented without any conventional SUSY particles.
\end{abstract}
\maketitle

There is an exciting possibility that the observed spectrum of elementary particles of
the standard model is obtainable from a suitable N=1, D=4 superstring massless modes of excitations.
If the supersymmetric group is $SO(6)\otimes SO(5)$, it can descend to $Z_3\otimes SU_C(3)\otimes
SU_L(2)\otimes U_Y(1)$ as a very special case ~\cite{deo}. Because the group $Z_3$ is enough 
to consider one of the three generations. $Z_3$ groups takes care of the other two generations.
The number of fermionic modes as observed today  in one generation is thirty due to the fifteen 
fermions. Similarly, including Higgs, the number of bosonic modes is twenty eight. If we
include two noninteracting scalar modes, the bosonic degrees of freedom will match
with those of the fermions. Therefore there could be a distinct supersymmetry scenario. 
With this aim in mind, we examine the case of a recently proposed new type of superstring 
in four dimensions~\cite{deo} where this is  realised.

 A world sheet $(\sigma, \tau)$ supersymmetric action is written as~\cite{deo}
\begin{equation}
S= -\frac{1}{2\pi}\int d^2\sigma\left [ \partial_{\alpha}X^{\mu}\partial^{\alpha}X_{\mu}
-i \bar{\psi}^{\mu,j}\rho^{\alpha}\partial_{\alpha}  \psi_{\mu,j}
+ i \bar{\phi}^{\mu,k}\rho^{\alpha}\partial_{\alpha}  \phi_{\mu,k}\right ].\label{a}
\end{equation}
$X^{\mu}$ are the bosonic coordinates. $\psi^{\mu,j}$, j=1,.,6 and $\phi^{\mu,k}$, k=7,.,11
are the Lorentz Majorana spinors in the bosonic representation of SO(3,1), with
\begin{eqnarray}
\rho^0 =
\left (
\begin{array}{cc}
0 & -i\\
i & 0\\
\end{array}
\right )
\end{eqnarray}
and
\begin{eqnarray}
\rho^1 =
\left (
\begin{array}{cc}
0 & i\\
i & 0\\
\end{array}
\right ).
\end{eqnarray}
The upper index in the action refers to a row and lower index to a column.
By counting the contributions of the fermions and bosons, the central charge of the 
action in equation (\ref{a}) is 26 like Nambu-Goto string but unlike a 
normal superstring. This is, because, the action also contains the light cone ghosts\\
\begin{equation}
S_{l.c.}=\frac{i}{2\pi}\;\int d^2\sigma \sum_{\mu =0,3}\left( \bar{\psi}^{\mu j}\rho^
{\alpha}\partial_{\alpha}\psi_{\mu j} - \bar{\phi}^{\mu k}\rho^
{\alpha}\partial_{\alpha}\phi_{\mu k}\right),\label{b}
\end{equation}
whose central charge adds up to eleven, like $\beta - \gamma$ superconformal ghost 
action. The action in equation (\ref{a}) is invariant under the supersymmetric transformations,

\begin{eqnarray}
\delta X^{\mu}& =&\bar{\epsilon}\;(\;e^j\;\psi^{\mu}_j - e^k\;\phi^{\mu}_k),\\
\delta\psi^{\mu,j}&=& - i\;e^j\;\rho^{\alpha}\;\partial_{\alpha}X^{\mu}\;\epsilon~~~~~~~~and  \\
\delta\phi^{\mu,k}&= &ie^k\;\rho^{\alpha}\partial_{\alpha}X^{\mu}\;\epsilon,
\end{eqnarray}
with $\epsilon$ as the constant anticommutating spinor. ${e}^j$ and
${e}^k$ are the two unit component c-number row vectors with the property that
${e}^j{e}_{j'}=\delta^j_{j'}$ and  ${e}^k{e}_{k'}=\delta^k_{k'}$ and it
follows that
${e}^j{e}_{j}$=6 and ${e}^k{e}_{k}$=5.
The commutator of two such transformations is a translation with coefficient
$a^{\alpha}=2i\bar{\epsilon}_1\rho^{\alpha}\epsilon_2$ provided
$\psi^{\mu,j}= e^j\Psi^{\mu}$ and
$\phi^{\mu,k}= e^k\Psi^{\mu}$ where
\begin{equation}
\Psi^{\mu}=e^j\;\psi^{\mu}_j - e^k\;\phi^{\mu}_k.
\end{equation}
and $\Psi^{\mu}$ is the superpartner of $X^{\mu}$ as can be checked. Writing out the 
2-d local supersymmetric action of Brink, Di Vecchia, Howe,Deser and Zumino~\cite{br},
we can write down  the Noether current in the gravitinoless gauge,
\begin{equation}
J_{\alpha}= \frac{\pi}{2e}\frac{\delta S}{\delta\bar{\chi}^{\alpha}}=\rho^{\beta}
\rho_{\alpha}\bar{\Psi}^{\mu}
\partial_{\beta}X_{\mu} = 0
\end{equation}
and in the simpler light cone co-ordinates
\begin{equation}
J_{\pm}=\partial_{\pm}X_{\mu}\Psi^{\mu}_{\pm}=0\label{J},
\end{equation}
and  the energy momentum tensor
\begin{equation}
T_{\pm\pm}=
\partial_{\pm}X^{\mu}\partial_{\pm}X_{\mu}+\frac{i}{ 2}{\psi}^{\mu j}_{\pm}\partial_{\pm}
 \psi_{\pm\mu,j }- \frac{i}{2}{\phi}_{\pm}^{\mu k}\partial_{\pm}\phi_{\pm\mu,k}.
\end{equation}
From action in equation (\ref{a}), commutation and anticommutation of the fields can be 
found and the string is quantised using Neveu-Schwarz and Ramond ~\cite{deo} boundary
conditions. The co-ordinate quanta, $\alpha^{\mu}_m$, satisfy 
\begin{equation}
[\alpha_m^{\mu}, \alpha_n^{\nu}]=m\delta_{m,-n}\eta^{\mu\nu}.
\end{equation}
In the NS sector, the quanta of $\psi$ and $\phi$ are respectively ~~$b_r^{\mu}$~~ and 
$b_r^{'\mu}$,~~ with r half integral and obey
\begin{equation}
\{ b^{\mu ,j}_r , b^{\nu,j'}_s\}=\eta^{\mu\nu}\delta^{j,j'}\delta_{r,-s} ~~~~~~~~~~and~~
~~~~~\{ b^{'\mu ,k}_r , b^{'\nu,k'}_s\}=-\eta^{\mu\nu}\delta^{k,k'}\delta_{r,-s}.
\end{equation}
Similarly the quanta d and d$'$ of R sector satisfy (m, n integers), the nonvanishing 
anticommutations
\begin{equation}
\{ d^{\mu ,j}_m , d^{\nu,j'}_n\}=\eta^{\mu\nu}\delta^{j,j'}\delta_{m,-n} ~~~~~~~~~and
~~~~~\{ d'^{\mu ,k}_m, d'^{\nu,k'}_n\}=-\eta^{\mu\nu}\delta^{k,k'}\delta_{m,-n}.
\end{equation}
The unusual negative signs are taken care of, by the phase factors,
\begin{equation}
b^{\prime\mu}_{-r,j}=-b^{\prime\dag\mu}_{r,j}\;\;\;\;\;\;\; and\;\;\;\;\;\;\;\;\;\;
d^{\prime\mu}_{-m,j}=-d^{\prime\dag\mu}_{m,j}.
\end{equation}
It will be useful to note ~~$ b_r^{\mu,j} = e^j C_r^{\mu},~~~
b_r^{\prime\mu,k} = e^k C_r^{\mu} ~~~~and~~~~C_r^{\mu} =
e^j b_{r,j}^{\mu} - e^k b_{r,k}^{\prime\mu}$.~~Similarly for the ~d, ~$d'$;\\
$D_m^{\mu} = e^j d_{m,j}^{\mu} - e^k d_{m,k}^{\prime\mu}$.

Virasoro generators~\cite{v} (dot implies summation of all qualifying indices and : :
for normal ordering)
\begin{eqnarray}
~~~~~~NS ~~~~~~~L_m&=&\frac{1}{2}\sum^{\infty}_{-\infty}:\alpha_{-n}\cdot\alpha_{m+n}: 
+\frac{1}{2}\sum_{r\in z+\frac{1}{2}}(r+\frac{1}{2}m): (b_{-r} \cdot b_{m+r} - 
b_{-r}' \cdot b_{m+r}'):
\nonumber\\
G_r &=&\frac{\sqrt{2}}{\pi}\int_{-\pi}^{\pi}d\sigma e^{ir\sigma}J_{+}=
\sum_{n=-\infty}^{\infty}\alpha_{-n}\cdot C_{n+r},\\
~~~~~~R~~~~~~~~~L_m&=&\frac{1}{2}\sum^{\infty}_{-\infty}:\alpha_{-n}\cdot\alpha_{m+n}: 
+\frac{1}{2}\sum^{\infty}_{n=-\infty}(n+\frac{1}{2}m): (d_{-n} \cdot d_{m+n} - d_{-n}'
\cdot d_{m+n}'):~,\\
F_r &=&\sum_{-\infty}^{\infty} \alpha_{-n}\cdot D_{n+r}.
\end{eqnarray}
By direct calculation, it is easily checked that the generators satisfy super Virasoro
algebra with anomaly term typical of central charge 26. To cancel these anomalies, 
we can use the known way of introducing the conformal ghosts b and c.\\
 
The BRST charge operator should also include 11 pairs of commuting~
$(\beta^{(j,k)}, \gamma^{(j,k)})$~ superconformal ghost system. Defining
\[
\beta_{r(m)}^{(j,k)} = e^{(j,k)}\beta_{r(m)}~~~~and ~~\gamma_{r(m)}^{(j,k)} 
= e^{(j,k)}\gamma_{r(m)}, \]  such that ~$\gamma_{r(m)}= e^j\gamma_{r(m),j}- 
e^k\gamma_{r(m),k}$ ~and  ~$\beta_{r(m)}= e^j\beta_{r(m),j}- e^k\beta_{r(m),k}$ ~with 
~$[ \beta_{r(m)}, \gamma_{s(n)}] =-\delta_{r+s,m+n}$.~  Since the conformal dimension 
of $\beta$ is~$\frac{3}{2}$~ and of ~$\gamma$~ is ~$-\frac{1}{2}$,~
\[ [L_m, \gamma_{s(n)}] =\left (-\frac{3}{2}m - s(n)\right )\gamma_{m+s(n)}~~~ and ~~~ 
[L_m, \beta_{r(n)}] = \left (\frac{1}{2}m - r(n) \right )\beta_{m+r(n)}.\]
In NS sector, the BRST charge is
\begin{eqnarray}
Q& =& \sum L_{-m}c_m -\frac{1}{2}\sum (m-n) : c_{-m}c_{-n}b_{m+n}:~ -~ c_0~a +Q',\\
Q'&=&\sum G_{-r}\gamma_r -\sum\gamma_{-r}\gamma_{-s}b_{r+s}
\end{eqnarray}
and  $Q^2 $=0 if a=1 is proved directly. The nilpotency proves the unitarity. Details can
be found in reference ~\cite{deo}.  As can be shortly shown, the physical state conditions 
remove all the ghosts from the Fock space,
\begin{eqnarray}
(L_o-1)|\phi>&=&0,\label{c}\\
(L_o-1)|\phi>_R&=&(F^2_o-1)|\phi>_R = (F_o+1)|\phi>_R = (F_o-1)|\phi>_R=0\\
L_m|\phi>&=&0,~~~~~for~~~m>0,\\
G_r|\phi>&=&0 ~~~~~for~~~~ r>0,~~~~~~~~~~~~~~and\label{d}\\
F_m|\phi>_R&=&0~~~~~ for~~~~~m>0.\label{e}
\end{eqnarray}
 We shall assume the Gupta- Bleuler condition~ $\alpha_{-1}^0|\phi>=0$.~
Since ~$[\alpha_{-1}^0 , L_{m+1}]|\phi>=0$, ~we get 
$\alpha_{m}^0|\phi>=0$. Further ~$[\alpha_{-1}^0 , G_{r+1}]|\phi>=0$~ leads to ~
$C_r^0|\phi>=0$ ~and~ $b^0_{r,j}|\phi>~=~e_{j}C^0_r|\phi>=0$;~ $b^{\prime 0}_{r,k}|\phi>=
e_{k}C^0_r|\phi>=0$. Similar analysis yields~
$d^{'0}_{r,k}|\phi>~=~d^0_{m,j}|\phi>=d^{\prime 0}_{m,k}|\phi>=0$.  
There is no Lorentz metric longitudinal ghost in the Fock space.\\
The half integral values of mass spectrum (NS), $\alpha' M^2= -1,0,1/2,1, 3/2$ 
are projected by G.S.O method. To show that the tachyons should not contribute to
the Hamiltonian, we construct the world sheet supersymmetric charge
\begin{eqnarray}
Q&=& \frac{i}{\pi}\int_0^{\pi} d\sigma\Psi^{\mu}\rho^{\alpha}\rho^0\partial_{\alpha}X_{\mu}\\
Q^{\dag}&=&- \frac{i}{\pi}\int_0^{\pi} d\sigma\Psi^{\mu}\rho^{\alpha\dag}\rho^0\partial_{\alpha}X_{\mu}
\end{eqnarray}
We that 
\begin{equation}
\sum_{\alpha}\{ Q^{\dag}, Q_{\alpha}\} = 2 H
\end{equation}
Thus 
\begin{equation}
\sum_{\alpha}|Q_{\alpha}|\phi_0>|^2=2<\phi_0|H|\phi_o>~\geq 0
\end{equation}
Obtaining a ghost free, anomaly free and unitary string, we calculate the zero mode string 
excitations leading to the Standard Model. This is easily understood as $<0|(L_o -1)^{-1}|0>$, 
the self energy is cancelled by\\
 $-<0|(F_0-1)^{-1}(F_0+1)^{-1}|0>_R$
of the fermion sector.The negative sign is due to the fermion loop.
The $B_{\mu}$, the hypercharge vector, is
simply given from the state
\begin{equation}
|\phi(p)> = \alpha_{-1}^{\mu}|0,p>B_{\mu}.
\end{equation}
$L_0$ condition ~(\ref{c}) tells that $B_{\mu}(p)$ is massless and $L_1$ condition gives
the Lorentz condition ~$p^{\mu}\,B_{\mu}=0$.

In NS sector,  the objects like ~~$b^{\mu}_{-\frac{1}{2},i} ~b^{\mu}_{-\frac{1}{2},j}$~~
are massless due to ~(\ref{c}). Dropping the suffix $-\frac{1}{2}$, we construct 
the traceless tensor, (i,j ~=~1,2,3), expressable in terms of vector fields $A^{\mu}_{ij}$,
\begin{eqnarray}
G^{\mu\nu}_{ij}&=&\left ( b^{\mu}_i b^{\nu}_j -  b^{\nu}_i b^{\mu}_j -
\frac{2}{3}\delta_{ij} b^{\mu}_l b^{\nu}_l\right )\\
&=& \partial^{\nu}A^{\mu}_{ij} - \partial^{\mu}A^{\nu}_{ij} +\left ( A^{\mu}_{il}
A^{\nu}_{jl}- A^{\nu}_{il}A^{\mu}_{jl}\right ).
\end{eqnarray}
The eight gluons are obtained by using the eight Gell-Mann ~$\lambda_l$~matrices.
\begin{equation}
V_l^{\mu}=(\lambda_l)_{ij}A^{\mu}_{ij}.\label{v1}
\end{equation}
Similarly for (i,j = 7,8), we get the $W^{\mu}$-mesons
\begin{eqnarray}
W^{\mu\nu}_{ij}&=&b_i^{'\mu} b_j^{'\nu} - b_i^{'\nu} b_j^{'\mu} - \delta_{ij}
b_l^{'\nu} b_l^{'\mu}\\
&=& \partial^{\nu}W^{\mu}_{ij} - \partial^{\mu}W^{\nu}_{ij} + \left ( W^{\mu}_{il}
W^{\nu}_{jl}- W^{\nu}_{il}W^{\mu}_{jl}\right ),
\end{eqnarray}
with
\begin{equation}
W^{\mu}_l = ({\tau_l})_{ij}W^{\mu}_{ij},\label{v2}
\end{equation}
where $\tau_l$, $l$=1,2,3 are the $2 \times 2$ isospin matrices.
Of the remaining $b^{\mu}_j$, j=4,5,6 and $b^{\prime\mu}_k$, k=9,10,11, we can form six 
Lorentz scalars ~~$\epsilon_{ijl}b^{\lambda}_j b^{\lambda}_l$ ~~and ~~$\epsilon_{ikm}
b^{'\lambda}_k b^{'\lambda}_m$.~~ Let us call them ~~$\phi_1, ...,\phi_6$.~~
The Higgs bosons are 
\begin{eqnarray}
\phi_H=\left (
\begin{array}{c}
\phi_1 + i\phi_2 \\
\phi_3 + i\phi_4 \\
\end{array}
\right ),\label{v3}
\end{eqnarray}
leaving behind a scalar 
\begin{equation}
\varphi_s = \phi_5 + i\phi_6.\label{v4}
\end{equation}

Let us consider the fermion sector. With $\Gamma^{\mu}$ as Dirac matrices and 
u(p) being the fermion wave function,  the states~(\ref{v1}),~(\ref{v2})~,(\ref{v3})~
and ~(\ref{v4}) with ~$\Gamma^{\mu}d_{-1,j,\mu}|0,p>u(p)$~ and 
~$\Gamma^{\mu}d'_{-1,k,\mu}|0,p>u(p)$~ are all massless states. Broadly j=1,..,6 
are the colour sector and k=7,..,11 are the electroweak sector. 
The current generator condition gives ~$\Gamma^{\mu}p_{\mu}u(p)=0$,~ the Dirac 
equation for each of them.

Recently one of us(BBD) has shown~\cite{deo} that using Wilson loop technique,
$SO(6)\otimes SO(5)$ can directly descend to the standard model reducing the rank 
by one and without breaking supersymmetry. In fact, it was proven that 
$SO(6)\otimes SO(5)\rightarrow Z_3\otimes SU_C(3)\otimes SU_L(2)\otimes U_Y(1)$ 
but we have only eleven Ramond Fermionic zero mass modes $d^{\mu}_{-1,j},~~
d^{'\mu}_{-1,k}$. We have to find a wayto put these fermions into electroweak 
and colour groups of one generation.

Let us drop the Dirac indices, the suffix -1, and prefix primes and supply a 
factor 2 such that
\begin{equation}
\left\{d_i, d_j\right\}=2\delta_{ij}
\end{equation}

Since SO(4) can come from either SO(6) or SO(5)~\cite{li}. We can choose any 
four of them and define new operators b's as follows
\begin{equation}
b_1=\frac{1}{2} (d_1 + i d_2),~~~~~b_1^*=\frac{1}{2} (d_1 - i d_2),~~~~~~~
b_2=\frac{1}{2} (d_3 + i d_4),~~~~and~~~~b_2^*=\frac{1}{2} (d_3 - i d_4).
\end{equation}          
They satisfy the algebra (i,j=1,2),
\begin{equation}
\left\{b_i, b_j\right\}=0,~~~~~\left\{b^*_i, b^*_j\right\}=0~~~and ~~~
\left\{b_i, b^*_j\right\}=  2\delta_{ij}
\end{equation}
This is the U(2) group, identified here as the isospin group. We now construct a table giving 
the assignment of all fermions(Table.1).
\begin{center}
\begin{table}[h]
\begin{tabular}{|c|c|c|c|c|}\hline\hline
Particles &  States                           & $I_3$ &  Y  &  Q\\ \hline 
$e_R$     & $\Gamma\cdot d^{'}_7|0>_{\alpha}$ &  0    & -2  & -1 \\ \hline
$ \left(
\begin{array}{c}
\nu_L \\ e_L \\ 
\end{array}  \right) 
$ 
&
$
\left ( \begin{array}{c}
 \frac{1}{\sqrt{2}} \Gamma\cdot (d'_8 + i d'_9)|0>_{\alpha}\\
 \frac{1}{\sqrt{2}} \Gamma\cdot (d'_{10} + id'_{11}) |0>_{\alpha} \\
\end{array} \right )     
$
&
$
\begin{array}{c}
1/2 \\
-1/2 \\
\end{array}
$  
&
$
\begin{array}{c}
-1 \\
-1 \\
\end{array}
$
&
$
\begin{array}{c}
0 \\
-1 \\ 
\end{array} $ \\ 
\hline
$u^{a,b,c}_R$  & $\Gamma\cdot d_{1,2,3}|0>_{\alpha}$ & 0 & 4/3 &2/3\\ \hline
$d^{a,b,c}_R$  & $\Gamma\cdot d_{4,5,6}|0>_{\alpha}$ & 0 & -2/3 &-1/3\\ \hline
$ \left (
\begin{array}{c}
u^{a,b,c}_L\\
d^{a,b,c}_L\\
\end{array}
\right )
$
&
$\left (
\begin{array}{c}
 \frac{1}{\sqrt{2}} \Gamma\cdot (d_{1} + i d_{4})|0>_{\alpha},
~~ \frac{1}{\sqrt{2}} \Gamma\cdot (d_{2} + i d_{5})|0>_{\alpha},
~~ \frac{1}{\sqrt{2}} \Gamma\cdot (d_{3} + i d_{6})|0>_{\alpha}, \\
~~ \frac{1}{\sqrt{2}}\Gamma\cdot (d^c_{1} +i d^c_{4})|0>_{\alpha} ,
~~ \frac{1}{\sqrt{2}}\Gamma\cdot (d^c_{2} + i d^c_{5})|0>_{\alpha}, 
~~ \frac{1}{\sqrt{2}}\Gamma\cdot (d^c_{3} + i d^c_{6})|0>_{\alpha}  \\
\end{array}
\right )
$
&
$
\begin{array}{c}
1/2\\
-1/2\\
\end{array}
$
&
$
\begin{array}{c}
1/3\\
1/3\\
\end{array}
$
&
$
\begin{array}{c}
2/3\\
-1/3\\
\end{array}$\\ 
\hline\hline
\end{tabular}
\caption{Particle Spectra, States and Quantum Numbers (SM)}
\end{table}
\end{center}
Once the weak isospins are fixed, the hypercharge can be calculated from the charge also.
There are fifteen fermions with 30 degrees of freedom matching the 30 bosonic degrees 
of freedom from above equations. 
We now proceed to construct the Hamiltonian which is supersymmetric and 
$SU_C(3)\otimes SU_L(2)\otimes U_Y(1)$ invariant following the ideas presented by one of
us~\cite{deo1}.
The vector fields denoted by $V^l_{\mu}$, l=1,..,8, are gluon fields. l=9 is the $U_Y(1)$
field and l=10,11,12  stand for W-mesons fields. We shall use the temporal gauge~\cite{c},
where $V^l_0$ =0. For each $V^l_i$, there are electric field strength $E_i^l$ and 
magnetic field strength  $B_i^l$. Following Nambu~\cite{n}, the combination 
\begin{equation}
F_i^l=\frac{1}{\sqrt{2}}\left [ E_i^l +  B_i^l\right ],
\end{equation}
satisfies the only nonvanishing equal time commutation relation
\begin{equation}
\left [ F_i^{\dagger l}(x),  F_j^{ m}(y)\right ] =i \delta^{lm}\epsilon_{ijk}
\partial^k(x-y).
\end{equation}
We then construct Wilson's loop line integrals to convert the ordinary derivatives acting
on fermion fields to respective gauge covariant derivatives.
The phase function for the colour is
\begin{equation}
U_C(x)= exp\left( ig\int_0^x\sum_{l=1}^{\infty}\lambda^l V_i^l dy_i\right).
\end{equation}
Denoting
\begin{equation}
Y(x)=g'\int_0^x B_i dy_i,
\end{equation}
the isospin phase functions are 
\begin{eqnarray}
U_Q(x)&=&exp\left(\frac{ig}{2}\int_0^x{\tau}\cdot{\bf W}_i dy_i -\frac{i}{6}Y(x)\right),\\
U(x)&=& exp\left( \frac{ig}{2}\int_0^x{\tau}\cdot{\bf W}_i dy_i -\frac{i}{2}Y(x)\right),\\
U_1(x)&=& exp\left(-\frac{2i}{3}Y(x)\right),\\
U_2(x)&=& exp\left(\frac{i}{3}Y(x)\right),\\
and \nonumber\\
U_R(x)&=& exp\left( iY(x) \right )
\end{eqnarray}
The $\psi^l$ will denote the fermions, $l$=1,..,6 refer to the coloured quark doublets. 
The sum of the products 
\begin{eqnarray}
\sum_{l=1}^6 F_i^{l\dagger}\cdot \psi^l &=&\sum_{l=1}^3\left (F_i^{*l}, F_i^{*l+3}\right )
\left(
\begin{array}{c}
\psi^l\\
\psi^{l+3}\\
\end{array}
\right )\\
&=&\sum_{l=1}^3\left ( F_i^{*l}, F_i^{*l+3} \right )U_Q~U_C
\left(
\begin{array}{c}
u^l_L\\
d^{l+3}_L\\
\end{array}
\right )
\end{eqnarray}
The singlet phased  colour quarks are
\begin{eqnarray}
\psi^l &=&U_1 U_C u^l_R,~~~~ l=7,8,9~~ and\\
\psi^l &=&U_2 U_C d^l_R,~~~~ l=10,11,12.
\end{eqnarray}
The Q=1,  singlet ~$\varphi_s$~ and the $e_R$  are phased as
\begin{equation}
\phi_s=U_R\varphi_s,~~~~~~~~~\psi_R = U_R e_R.
\end{equation}
The Higgs doublets $\phi$ and the lepton doublet $\psi^l$  are phased like
\begin{equation}
\phi = U\phi_H~~~~~~~~~~~~ \psi_L=U
\left(
\begin{array}{c}
\nu_e\\
e^-\\
\end{array}
\right ).
\end{equation}
In writing down the supersymmetric charge Q, we shall use $2\times 2$ matrices, $\sigma^{\mu},
~~\mu = 0,1,2,3;~~ \sigma^0 =I$ and ${\bf \sigma}$'s are the three Pauli spin matrices,
\begin{equation}
Q=\int d^3 x \left [ \sum_{l=1}^{12}\left ( {\mathbf{\sigma}}\cdot {\bf F}^{\dagger l}
\psi^l(x) \right )+
\sigma^{\mu}\partial_{\mu}\phi^{\dagger}(x)\cdot\psi_L(x) +\sigma^{\mu}\partial_{\mu}
\phi_s^{\dagger}(x)\cdot \psi_L(x) + \mathcal{W}~\psi^1 \right ].
\end{equation}
where $\mathcal{W} = \sqrt{\lambda}~ [\phi^{\dagger}\phi - \upsilon^2~]$ is the electroweak 
symmetry breaking term with $\sigma=\sqrt{2}\upsilon = 246 ~~GeV$. We use the usual 
commutators and anticommutators at equal times and the $\sigma_i\sigma_j=
\delta_{ij} + i \epsilon_{ijk}\sigma_k$ and obtain
\begin{equation}
\left \{Q_{\alpha}^{\dagger},Q_{\beta}\right \}= \left( \sigma^{\mu} P_{\mu}\right )_
{\alpha\beta},
\end{equation}
$P_0$ is the Hamiltonian H and P is the total momentum. We get the
$SU_C(3)\otimes SU_L(2)\otimes U_Y(1)$ invariant Hamiltonian as
\begin{eqnarray}
H&=&\int d^3 x~  [~ \sum_{l=1}^{12}\frac{1}{2} \left ({\bf E}^{l2} + {\bf B}^{l2} +i 
\psi^{l\dagger}(x){\mathbf \sigma}\cdot {\mathbf \nabla}\psi^l(x) \right )\nonumber\\
&+& \pi^{\dagger}_H(x) \pi_H(x) + ({\mathbf\nabla}\phi^{\dagger})\cdot {\mathbf\nabla}\phi + 
i\psi^{\dagger}_L(x){\mathbf\sigma}\cdot {\mathbf\nabla}\psi_L(x)
+ i\psi^{\dagger}_R(x){\mathbf\sigma}\cdot {\mathbf\nabla}\psi_R(x) +\mathcal{W}^2~ ]
\nonumber\\
&+& \int d^3 x~\left [~\pi^{\dagger}_s(x) \pi_s(x)+ ({\mathbf {\nabla}}\phi_s^{\dagger})\cdot 
{\mathbf{\nabla}}\phi_s\right ]
\end{eqnarray}
and  similarly lengthy expression for P (see reference [3]). The last two terms involving 
Y=2 meson is absent in the standard model. But the Hamilton's equation of motion for 
the extra meson, obtained from the total Hamiltonian, is
\begin{equation}
\left ( \frac{\partial^2}{\partial t^2} -{\bf{\nabla}}^2 \right )\phi_s =0.
\end{equation}
and decouples from other standard model particles.

Supersymmetric transformations exhibiting the superpartners are
\begin{eqnarray}
\delta\psi^l &=& {\bf \sigma} \cdot{\bf  F}^l~\epsilon + \delta_{l1} \mathcal{W}~\epsilon ,\\
\delta {\bf F}_i^l &=& -i ~\bar{\epsilon} ~({\bf\sigma}\times{\bf\nabla})_i~
\psi^l ,\\
\delta \psi^l_L &=& \sigma^{\mu}\partial_{\mu}\phi~\epsilon ,\\
\delta \phi &=&-i\bar{\epsilon}\psi_L ,\\
\delta \psi_R &=& \sigma^{\mu}\partial_{\mu}\phi_s~\epsilon ,\\
\delta \phi_s &=&~-i~\bar{\epsilon}~\psi_L .
\end{eqnarray}
The only particle, not in the standard model, $\phi_s$, is noninteracting and decoupled.

In summary, we wish to emphasise that the proposed superstring excitations of zero 
mass fall into the standard model particles in a very natural way. The bosons are 
themselves, the superpartners of the fermions and vice versa. It is not surprising that 
the  exotic SUSYs are not observed in the electroweak scale and the present formulation
provides a justification for their absence.\\


\begin{thebibliography}{99}
\bibitem{deo} B. B. Deo, ~Phys. Lett. {\bf B 557}(2003) 115 ,\\ 
B. B. Deo and L. Maharana, ~{\it Derivation of 
Einstein's equation from a new type of superstring in four\\ dimensions} hep-th/0212004.
B. B. Deo, {\it Supersymmetric Standard Model from String Theory} hep-th/0301041
\bibitem{br} L. Brink, P. Di Vecchia and P. Howe, Phys. Lett. {\bf 65B}(1976) 471;
S. Deser and B. Zumino, Phys. Lett. {\bf 65B}(1976) 369.
\bibitem{v}  M.A. Virasoro, Phys. Rev. {\bf D1}(1970) 2933,
\bibitem{li} Linf-Fong Li, Phys. Rev. {\bf D9}, (1974) 1723
\bibitem{deo1} B. B. Deo, Mod. Phys. Lett.  {\bf A13}(1998) 2971,
\bibitem{c} N. H. Christ and T. D. Lee, Phys. Rev. {\bf D22}(1980) 939,
\bibitem{n} Y. Nambu, {\it Supersymmetry and Quasisupersymmetry}, E.F. Reprints(unpublished)
1991,
\end{thebibliography}
\end{document}